\documentstyle[emlines,emlines2,bezier]{article}

\begin{document}

\vspace*{2cm}
\begin{center}
 \Huge\bf
Large fluctuations of time and
change of space-time signature
\vspace*{0.25in}

\large

Alexander K.\ Guts, Marina S.\ Shapovalova
\vspace*{0.15in}

\normalsize

Department of Mathematics, Omsk State University \\
644077 Omsk-77 RUSSIA
\\
\vspace*{0.5cm}
E-mail: guts@univer.omsk.su  \\
shapovalova$_{-}{\rm m}$@mail.ru

\vspace*{0.5cm}
December 22, 2000\\
\vspace{.5in}
ABSTRACT
\end{center}

\vspace{.3in}
We consider five-dimensional cylindric spacetime $V^5$ with foliation of
codimension ~1. The leaves of this foliation are four-dimensional
 "parallel" universes. The metric of five-dimensional spacetime and induced metrics
 of four-dimensional universes are flat.
The "large" fluctuations of the 5-metric are studied. These fluctuations
 depend only on the coordinate $x^0$, and under these fluctuations the
 curvature of $V^5$ is not zero.  The contribution  of the fluctuations in the Feynman path
 integral over five-dimensional trajectories
doesn't change the amplitude of the probability of the real
physical four-dimensional universe. Moreover the large fluctuations of
5-metric  $G_{AB}$ are large fluctuations for physical four-dimensional
universe $V^4$  and change  signature of $V^4$.
The change of the signature from  $<+--->$ to $<---->$ and inversely occurs
in the all 3-dimensional space simultaneously (in absolute time) and can
take arbitrarily large period of time.

\newpage

\setcounter{page}{1}

\def\R{{{\rm I} \! {\rm R}}}

\section{Introduction}

It is well-known that spacetime at the Plank's scale can have strong
fluctuations
of metric. Spacetime at this scale is a  "quantum foam" with a
high curvature and a veriable topology.

The aim of this work is to construct the metric fluctuations which can exist
not only at the Plank's scale but at any distance. Such fluctuations are called
"large" \cite{1,2}.

The Kaluza-Klein theory will be used. We find the large fluctuations of
5-metric  $G_{AB}$ which are large one for physical four-dimensional
global hyperbolic universe $V^4$  and change  signature of $V^4$.
The change of the signature from  $<+--->$ to $<---->$ and inversely occurs
in the all 3-dimensional space simultaneously (in absolute time) and can
take arbitrarily large period of time.

\section{Basic spacetime}

As an example we consider the flat 5-dimensional spacetime with metric
$$
dI^2=G_{AB}dx^{A}dx^{B}=
$$
$$
= (dx^{0})^{2} -\beta(x^{0}-1)^{2}(dx^{1})^{2}-(dx^{2})^{2}
-(dx^{3})^{2}-(dx^{4})^{2}, \eqno(1)
$$
  where  $x^0, x^1$ is a polar coordinates, $x^0$ is a radius and $x^1$
is an angle (see fig.1). Here $\beta$ is a constant from the interval
$(0,1)$ and $x^0>1$. It is evident that this metric has a signature $<+---->$ and
$$
 V^{5}=\R^{1}\times S^{1}\times \R^{3}
 $$
 The metric (1) is flat.

\begin{figure}[h]

\begin{center}
\special{em:linewidth 0.4pt}
\unitlength 0.7mm
\linethickness{0.4pt}
\begin{picture}(146.67,144.00)
\emline{54.00}{117.74}{1}{56.56}{117.47}{2}
\emline{56.56}{117.47}{3}{59.01}{116.68}{4}
\emline{59.01}{116.68}{5}{61.25}{115.40}{6}
\emline{61.25}{115.40}{7}{63.17}{113.69}{8}
\emline{63.17}{113.69}{9}{64.70}{111.61}{10}
\emline{64.70}{111.61}{11}{65.77}{109.27}{12}
\emline{65.77}{109.27}{13}{66.32}{106.75}{14}
\emline{66.32}{106.75}{15}{66.35}{104.18}{16}
\emline{66.35}{104.18}{17}{65.85}{101.65}{18}
\emline{65.85}{101.65}{19}{64.83}{99.28}{20}
\emline{64.83}{99.28}{21}{63.35}{97.18}{22}
\emline{63.35}{97.18}{23}{61.46}{95.42}{24}
\emline{61.46}{95.42}{25}{59.25}{94.09}{26}
\emline{59.25}{94.09}{27}{56.82}{93.25}{28}
\emline{56.82}{93.25}{29}{54.26}{92.93}{30}
\emline{54.26}{92.93}{31}{51.69}{93.14}{32}
\emline{51.69}{93.14}{33}{49.23}{93.88}{34}
\emline{49.23}{93.88}{35}{46.96}{95.12}{36}
\emline{46.96}{95.12}{37}{45.01}{96.79}{38}
\emline{45.01}{96.79}{39}{43.44}{98.83}{40}
\emline{43.44}{98.83}{41}{42.32}{101.15}{42}
\emline{42.32}{101.15}{43}{41.71}{103.66}{44}
\emline{41.71}{103.66}{45}{41.63}{106.23}{46}
\emline{41.63}{106.23}{47}{42.08}{108.77}{48}
\emline{42.08}{108.77}{49}{43.05}{111.15}{50}
\emline{43.05}{111.15}{51}{44.48}{113.29}{52}
\emline{44.48}{113.29}{53}{46.33}{115.09}{54}
\emline{46.33}{115.09}{55}{48.51}{116.46}{56}
\emline{48.51}{116.46}{57}{50.93}{117.35}{58}
\emline{50.93}{117.35}{59}{54.00}{117.74}{60}
\emline{40.67}{102.00}{61}{41.36}{99.85}{62}
\emline{41.36}{99.85}{63}{42.17}{97.87}{64}
\emline{42.17}{97.87}{65}{43.10}{96.06}{66}
\emline{43.10}{96.06}{67}{44.13}{94.43}{68}
\emline{44.13}{94.43}{69}{45.28}{92.97}{70}
\emline{45.28}{92.97}{71}{46.54}{91.69}{72}
\emline{46.54}{91.69}{73}{47.92}{90.59}{74}
\emline{47.92}{90.59}{75}{49.41}{89.66}{76}
\emline{49.41}{89.66}{77}{51.01}{88.90}{78}
\emline{51.01}{88.90}{79}{52.72}{88.32}{80}
\emline{52.72}{88.32}{81}{54.55}{87.91}{82}
\emline{54.55}{87.91}{83}{56.49}{87.68}{84}
\emline{56.49}{87.68}{85}{58.54}{87.63}{86}
\emline{58.54}{87.63}{87}{60.70}{87.74}{88}
\emline{60.70}{87.74}{89}{62.98}{88.04}{90}
\emline{62.98}{88.04}{91}{65.37}{88.51}{92}
\emline{65.37}{88.51}{93}{67.88}{89.15}{94}
\emline{67.88}{89.15}{95}{70.49}{89.97}{96}
\emline{70.49}{89.97}{97}{73.22}{90.96}{98}
\emline{73.22}{90.96}{99}{76.07}{92.13}{100}
\emline{76.07}{92.13}{101}{79.02}{93.47}{102}
\emline{79.02}{93.47}{103}{82.09}{94.99}{104}
\emline{82.09}{94.99}{105}{85.27}{96.68}{106}
\emline{85.27}{96.68}{107}{88.57}{98.54}{108}
\emline{88.57}{98.54}{109}{91.97}{100.59}{110}
\emline{91.97}{100.59}{111}{95.49}{102.80}{112}
\emline{95.49}{102.80}{113}{99.33}{105.33}{114}
\emline{54.00}{91.67}{115}{56.80}{92.03}{116}
\emline{56.80}{92.03}{117}{59.41}{92.51}{118}
\emline{59.41}{92.51}{119}{61.84}{93.10}{120}
\emline{61.84}{93.10}{121}{64.08}{93.80}{122}
\emline{64.08}{93.80}{123}{66.14}{94.62}{124}
\emline{66.14}{94.62}{125}{68.02}{95.54}{126}
\emline{68.02}{95.54}{127}{69.72}{96.58}{128}
\emline{69.72}{96.58}{129}{71.23}{97.73}{130}
\emline{71.23}{97.73}{131}{72.55}{98.99}{132}
\emline{72.55}{98.99}{133}{73.70}{100.37}{134}
\emline{73.70}{100.37}{135}{74.65}{101.85}{136}
\emline{74.65}{101.85}{137}{75.43}{103.45}{138}
\emline{75.43}{103.45}{139}{76.02}{105.16}{140}
\emline{76.02}{105.16}{141}{76.43}{106.98}{142}
\emline{76.43}{106.98}{143}{76.65}{108.92}{144}
\emline{76.65}{108.92}{145}{76.69}{110.96}{146}
\emline{76.69}{110.96}{147}{76.55}{113.12}{148}
\emline{76.55}{113.12}{149}{76.22}{115.39}{150}
\emline{76.22}{115.39}{151}{75.71}{117.77}{152}
\emline{75.71}{117.77}{153}{75.01}{120.27}{154}
\emline{75.01}{120.27}{155}{74.14}{122.87}{156}
\emline{74.14}{122.87}{157}{73.07}{125.59}{158}
\emline{73.07}{125.59}{159}{71.83}{128.42}{160}
\emline{71.83}{128.42}{161}{70.40}{131.36}{162}
\emline{70.40}{131.36}{163}{68.78}{134.41}{164}
\emline{68.78}{134.41}{165}{66.99}{137.58}{166}
\emline{66.99}{137.58}{167}{65.00}{140.86}{168}
\emline{65.00}{140.86}{169}{63.00}{144.00}{170}
\emline{67.67}{102.67}{171}{67.88}{104.87}{172}
\emline{67.88}{104.87}{173}{67.92}{106.97}{174}
\emline{67.92}{106.97}{175}{67.79}{108.96}{176}
\emline{67.79}{108.96}{177}{67.47}{110.86}{178}
\emline{67.47}{110.86}{179}{66.98}{112.64}{180}
\emline{66.98}{112.64}{181}{66.32}{114.33}{182}
\emline{66.32}{114.33}{183}{65.48}{115.91}{184}
\emline{65.48}{115.91}{185}{64.46}{117.38}{186}
\emline{64.46}{117.38}{187}{63.26}{118.75}{188}
\emline{63.26}{118.75}{189}{61.89}{120.02}{190}
\emline{61.89}{120.02}{191}{60.35}{121.18}{192}
\emline{60.35}{121.18}{193}{58.62}{122.24}{194}
\emline{58.62}{122.24}{195}{56.72}{123.20}{196}
\emline{56.72}{123.20}{197}{54.65}{124.05}{198}
\emline{54.65}{124.05}{199}{52.40}{124.79}{200}
\emline{52.40}{124.79}{201}{49.97}{125.44}{202}
\emline{49.97}{125.44}{203}{47.36}{125.98}{204}
\emline{47.36}{125.98}{205}{44.58}{126.41}{206}
\emline{44.58}{126.41}{207}{41.62}{126.74}{208}
\emline{41.62}{126.74}{209}{38.49}{126.97}{210}
\emline{38.49}{126.97}{211}{35.18}{127.09}{212}
\emline{35.18}{127.09}{213}{31.69}{127.11}{214}
\emline{31.69}{127.11}{215}{28.03}{127.02}{216}
\emline{28.03}{127.02}{217}{24.19}{126.83}{218}
\emline{24.19}{126.83}{219}{20.18}{126.54}{220}
\emline{20.18}{126.54}{221}{14.67}{126.00}{222}
\emline{53.67}{119.33}{223}{51.31}{118.98}{224}
\emline{51.31}{118.98}{225}{49.08}{118.49}{226}
\emline{49.08}{118.49}{227}{46.99}{117.85}{228}
\emline{46.99}{117.85}{229}{45.03}{117.08}{230}
\emline{45.03}{117.08}{231}{43.20}{116.17}{232}
\emline{43.20}{116.17}{233}{41.51}{115.12}{234}
\emline{41.51}{115.12}{235}{39.95}{113.92}{236}
\emline{39.95}{113.92}{237}{38.52}{112.59}{238}
\emline{38.52}{112.59}{239}{37.22}{111.12}{240}
\emline{37.22}{111.12}{241}{36.06}{109.51}{242}
\emline{36.06}{109.51}{243}{35.03}{107.76}{244}
\emline{35.03}{107.76}{245}{34.13}{105.87}{246}
\emline{34.13}{105.87}{247}{33.36}{103.84}{248}
\emline{33.36}{103.84}{249}{32.73}{101.67}{250}
\emline{32.73}{101.67}{251}{32.23}{99.36}{252}
\emline{32.23}{99.36}{253}{31.86}{96.91}{254}
\emline{31.86}{96.91}{255}{31.63}{94.32}{256}
\emline{31.63}{94.32}{257}{31.53}{91.59}{258}
\emline{31.53}{91.59}{259}{31.56}{88.72}{260}
\emline{31.56}{88.72}{261}{31.72}{85.72}{262}
\emline{31.72}{85.72}{263}{32.02}{82.57}{264}
\emline{32.02}{82.57}{265}{32.45}{79.28}{266}
\emline{32.45}{79.28}{267}{33.01}{75.85}{268}
\emline{33.01}{75.85}{269}{33.70}{72.29}{270}
\emline{33.70}{72.29}{271}{35.33}{65.33}{272}
\put(72.67,73.00){\makebox(0,0)[cc]{$V^5$}}
\put(82.67,130.33){\makebox(0,0)[cc]{$V^4$}}
\put(75.00,121.33){\vector(-1,-2){0.2}}
\emline{78.00}{128.00}{273}{75.00}{121.33}{274}
\put(106.67,105.00){\vector(1,0){0.2}}
\emline{54.00}{105.00}{275}{106.67}{105.00}{276}
\emline{54.33}{105.33}{277}{76.67}{114.00}{278}
\emline{76.67}{113.67}{279}{77.94}{112.15}{280}
\emline{77.94}{112.15}{281}{78.75}{110.25}{282}
\emline{78.75}{110.25}{283}{79.10}{107.98}{284}
\emline{79.10}{107.98}{285}{79.00}{105.33}{286}
\put(71.67,115.67){\makebox(0,0)[cc]{$x^0$}}
\put(82.67,110.33){\makebox(0,0)[cc]{$x^1$}}
\put(87.00,114.33){\makebox(0,0)[cc]{$(x^0,x^1)$}}
\end{picture}

\vspace{-4cm}
\caption{Spacetime $V^5$ with leaf $V^4$}
\end{center}
\end{figure}
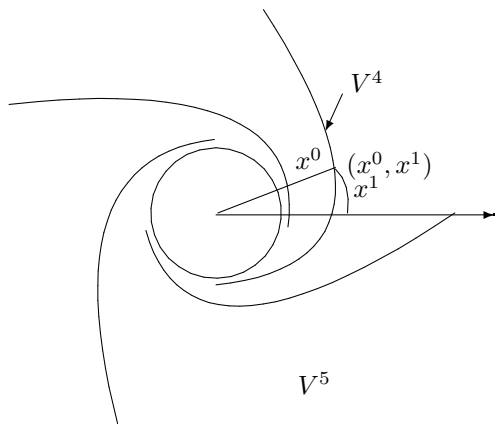

\vspace{1cm}

The physical 4-dimensional universes are defined
by equation
$$
x^{0}=1+\alpha \exp(x^{1}),\eqno(2)
$$
where $0<\alpha<+\infty$ is constant. Each $\alpha$ gives some
leaf $V^{4}_\alpha$.
The equation (2) corresponds to a spiral in the plane $(x^{0},x^{1})$ (fig.1)

Consider the imbedding $f:V^5\to \R^6$ of $V^5$ into $\R^6$
$$
\left\{
\begin{array}{ll}
  u^{0}=x^{0}-1 \\
  u^{1}=\gamma \cos(x^{1})  \\
  u^{2}=\gamma \sin(x^{1})  \\
  u^{3}=x^{2}   \\
  u^{4}=x^{3}   \\
  u^{5}=x^{4},
\end{array}\right.
$$
Then $f(V^5)$ is cylinder in $\R^6$
$$
 f(V^5)=\{(u^{0},u^{1},u^{2},u^{3},
u^{4},u^{5})
\in  R^{6}: u^{0}>0\  \& \
(u^{1})^{2}+(u^{2})^{2}=\gamma^{2}\}
$$
where $\gamma$ is some  constant, radius of cylinder (fig.2).

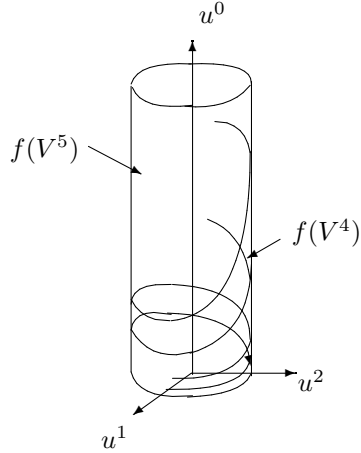
\begin{figure}[h]
\begin{center}
\special{em:linewidth 0.4pt}
\unitlength 0.5mm
\linethickness{0.4pt}
\begin{picture}(86.00,148.00)
\emline{38.00}{60.00}{1}{38.61}{58.37}{2}
\emline{38.61}{58.37}{3}{39.55}{56.98}{4}
\emline{39.55}{56.98}{5}{40.80}{55.82}{6}
\emline{40.80}{55.82}{7}{42.37}{54.89}{8}
\emline{42.37}{54.89}{9}{44.25}{54.19}{10}
\emline{44.25}{54.19}{11}{46.46}{53.73}{12}
\emline{46.46}{53.73}{13}{48.99}{53.50}{14}
\emline{48.99}{53.50}{15}{54.33}{53.67}{16}
\emline{53.00}{53.33}{17}{56.39}{53.56}{18}
\emline{56.39}{53.56}{19}{59.41}{53.93}{20}
\emline{59.41}{53.93}{21}{62.06}{54.45}{22}
\emline{62.06}{54.45}{23}{64.34}{55.10}{24}
\emline{64.34}{55.10}{25}{66.25}{55.90}{26}
\emline{66.25}{55.90}{27}{67.80}{56.84}{28}
\emline{67.80}{56.84}{29}{68.97}{57.92}{30}
\emline{68.97}{57.92}{31}{70.00}{59.67}{32}
\emline{38.33}{135.67}{33}{39.17}{134.21}{34}
\emline{39.17}{134.21}{35}{40.33}{133.00}{36}
\emline{40.33}{133.00}{37}{41.83}{132.04}{38}
\emline{41.83}{132.04}{39}{43.67}{131.33}{40}
\emline{43.67}{131.33}{41}{45.83}{130.88}{42}
\emline{45.83}{130.88}{43}{48.33}{130.67}{44}
\emline{48.33}{130.67}{45}{51.17}{130.71}{46}
\emline{51.17}{130.71}{47}{54.33}{131.00}{48}
\emline{52.33}{130.67}{49}{55.17}{130.54}{50}
\emline{55.17}{130.54}{51}{57.79}{130.63}{52}
\emline{57.79}{130.63}{53}{60.21}{130.93}{54}
\emline{60.21}{130.93}{55}{62.42}{131.45}{56}
\emline{62.42}{131.45}{57}{64.42}{132.19}{58}
\emline{64.42}{132.19}{59}{66.21}{133.15}{60}
\emline{66.21}{133.15}{61}{67.80}{134.32}{62}
\emline{67.80}{134.32}{63}{69.67}{136.33}{64}
\emline{38.00}{136.67}{65}{38.69}{138.06}{66}
\emline{38.69}{138.06}{67}{39.75}{139.25}{68}
\emline{39.75}{139.25}{69}{41.19}{140.23}{70}
\emline{41.19}{140.23}{71}{43.00}{141.00}{72}
\emline{43.00}{141.00}{73}{45.19}{141.56}{74}
\emline{45.19}{141.56}{75}{47.75}{141.92}{76}
\emline{47.75}{141.92}{77}{50.69}{142.06}{78}
\emline{50.69}{142.06}{79}{54.00}{142.00}{80}
\emline{53.33}{142.00}{81}{56.82}{142.07}{82}
\emline{56.82}{142.07}{83}{59.91}{141.97}{84}
\emline{59.91}{141.97}{85}{62.58}{141.73}{86}
\emline{62.58}{141.73}{87}{64.84}{141.32}{88}
\emline{64.84}{141.32}{89}{66.69}{140.76}{90}
\emline{66.69}{140.76}{91}{68.13}{140.03}{92}
\emline{68.13}{140.03}{93}{69.17}{139.16}{94}
\emline{69.17}{139.16}{95}{69.79}{138.12}{96}
\emline{69.79}{138.12}{97}{70.00}{136.67}{98}
\emline{38.00}{71.33}{99}{39.29}{69.44}{100}
\emline{39.29}{69.44}{101}{40.72}{67.86}{102}
\emline{40.72}{67.86}{103}{42.29}{66.59}{104}
\emline{42.29}{66.59}{105}{44.02}{65.63}{106}
\emline{44.02}{65.63}{107}{45.89}{64.98}{108}
\emline{45.89}{64.98}{109}{47.91}{64.64}{110}
\emline{47.91}{64.64}{111}{50.07}{64.61}{112}
\emline{50.07}{64.61}{113}{54.33}{65.33}{114}
\emline{53.67}{65.00}{115}{55.81}{65.72}{116}
\emline{55.81}{65.72}{117}{57.81}{66.57}{118}
\emline{57.81}{66.57}{119}{59.65}{67.57}{120}
\emline{59.65}{67.57}{121}{61.35}{68.70}{122}
\emline{61.35}{68.70}{123}{62.89}{69.97}{124}
\emline{62.89}{69.97}{125}{64.28}{71.38}{126}
\emline{64.28}{71.38}{127}{65.52}{72.92}{128}
\emline{65.52}{72.92}{129}{66.61}{74.61}{130}
\emline{66.61}{74.61}{131}{67.55}{76.43}{132}
\emline{67.55}{76.43}{133}{68.33}{78.39}{134}
\emline{68.33}{78.39}{135}{68.97}{80.49}{136}
\emline{68.97}{80.49}{137}{69.45}{82.73}{138}
\emline{69.45}{82.73}{139}{69.79}{85.11}{140}
\emline{69.79}{85.11}{141}{70.00}{88.67}{142}
\emline{38.00}{78.00}{143}{38.26}{79.62}{144}
\emline{38.26}{79.62}{145}{39.04}{80.96}{146}
\emline{39.04}{80.96}{147}{40.34}{81.99}{148}
\emline{40.34}{81.99}{149}{42.17}{82.74}{150}
\emline{42.17}{82.74}{151}{44.51}{83.19}{152}
\emline{44.51}{83.19}{153}{48.67}{83.33}{154}
\emline{69.33}{62.33}{155}{68.75}{61.04}{156}
\emline{68.75}{61.04}{157}{67.87}{59.88}{158}
\emline{67.87}{59.88}{159}{66.70}{58.85}{160}
\emline{66.70}{58.85}{161}{65.23}{57.95}{162}
\emline{65.23}{57.95}{163}{63.46}{57.19}{164}
\emline{63.46}{57.19}{165}{61.39}{56.55}{166}
\emline{61.39}{56.55}{167}{59.03}{56.05}{168}
\emline{59.03}{56.05}{169}{56.37}{55.68}{170}
\emline{56.37}{55.68}{171}{53.41}{55.44}{172}
\emline{53.41}{55.44}{173}{47.33}{55.33}{174}
\emline{69.67}{84.67}{175}{69.21}{87.25}{176}
\emline{69.21}{87.25}{177}{68.56}{89.62}{178}
\emline{68.56}{89.62}{179}{67.73}{91.79}{180}
\emline{67.73}{91.79}{181}{66.71}{93.75}{182}
\emline{66.71}{93.75}{183}{65.49}{95.51}{184}
\emline{65.49}{95.51}{185}{64.09}{97.07}{186}
\emline{64.09}{97.07}{187}{62.50}{98.42}{188}
\emline{62.50}{98.42}{189}{60.72}{99.56}{190}
\emline{60.72}{99.56}{191}{58.33}{100.67}{192}
\emline{70.00}{137.33}{193}{70.00}{59.00}{194}
\emline{48.67}{73.67}{195}{50.49}{73.92}{196}
\emline{50.49}{73.92}{197}{52.23}{74.34}{198}
\emline{52.23}{74.34}{199}{53.88}{74.93}{200}
\emline{53.88}{74.93}{201}{55.46}{75.67}{202}
\emline{55.46}{75.67}{203}{56.95}{76.57}{204}
\emline{56.95}{76.57}{205}{58.36}{77.64}{206}
\emline{58.36}{77.64}{207}{59.68}{78.86}{208}
\emline{59.68}{78.86}{209}{60.93}{80.25}{210}
\emline{60.93}{80.25}{211}{62.09}{81.80}{212}
\emline{62.09}{81.80}{213}{63.17}{83.51}{214}
\emline{63.17}{83.51}{215}{64.16}{85.38}{216}
\emline{64.16}{85.38}{217}{65.07}{87.42}{218}
\emline{65.07}{87.42}{219}{65.90}{89.61}{220}
\emline{65.90}{89.61}{221}{66.65}{91.97}{222}
\emline{66.65}{91.97}{223}{67.32}{94.49}{224}
\emline{67.32}{94.49}{225}{67.90}{97.17}{226}
\emline{67.90}{97.17}{227}{68.40}{100.01}{228}
\emline{68.40}{100.01}{229}{68.81}{103.01}{230}
\emline{68.81}{103.01}{231}{69.15}{106.17}{232}
\emline{69.15}{106.17}{233}{69.40}{109.50}{234}
\emline{69.40}{109.50}{235}{69.57}{112.98}{236}
\emline{69.57}{112.98}{237}{69.67}{118.67}{238}
\emline{69.67}{118.33}{239}{69.09}{120.73}{240}
\emline{69.09}{120.73}{241}{68.18}{122.73}{242}
\emline{68.18}{122.73}{243}{66.95}{124.31}{244}
\emline{66.95}{124.31}{245}{65.40}{125.49}{246}
\emline{65.40}{125.49}{247}{63.51}{126.27}{248}
\emline{63.51}{126.27}{249}{60.33}{126.67}{250}
\put(54.33,148.00){\vector(0,1){0.2}}
\emline{54.33}{59.33}{251}{54.33}{148.00}{252}
\put(81.67,59.67){\vector(1,0){0.2}}
\emline{54.00}{59.67}{253}{81.67}{59.67}{254}
\put(38.67,48.67){\vector(-4,-3){0.2}}
\emline{54.00}{59.67}{255}{38.67}{48.67}{256}
\emline{38.00}{136.00}{257}{38.00}{60.00}{258}
\emline{38.00}{79.33}{259}{39.21}{77.39}{260}
\emline{39.21}{77.39}{261}{40.70}{75.86}{262}
\emline{40.70}{75.86}{263}{42.46}{74.72}{264}
\emline{42.46}{74.72}{265}{44.50}{73.99}{266}
\emline{44.50}{73.99}{267}{48.33}{73.67}{268}
\emline{38.00}{71.00}{269}{38.40}{72.68}{270}
\emline{38.40}{72.68}{271}{39.20}{74.00}{272}
\emline{39.20}{74.00}{273}{40.40}{74.97}{274}
\emline{40.40}{74.97}{275}{42.01}{75.57}{276}
\emline{42.01}{75.57}{277}{44.03}{75.81}{278}
\emline{44.03}{75.81}{279}{48.67}{75.33}{280}
\put(69.67,62.33){\vector(1,-4){0.2}}
\emline{47.67}{75.67}{281}{50.65}{75.50}{282}
\emline{50.65}{75.50}{283}{53.41}{75.21}{284}
\emline{53.41}{75.21}{285}{55.96}{74.79}{286}
\emline{55.96}{74.79}{287}{58.29}{74.24}{288}
\emline{58.29}{74.24}{289}{60.40}{73.55}{290}
\emline{60.40}{73.55}{291}{62.30}{72.74}{292}
\emline{62.30}{72.74}{293}{63.99}{71.80}{294}
\emline{63.99}{71.80}{295}{65.45}{70.73}{296}
\emline{65.45}{70.73}{297}{66.70}{69.53}{298}
\emline{66.70}{69.53}{299}{67.74}{68.20}{300}
\emline{67.74}{68.20}{301}{68.55}{66.74}{302}
\emline{68.55}{66.74}{303}{69.15}{65.15}{304}
\emline{69.15}{65.15}{305}{69.67}{62.33}{306}
\emline{49.00}{83.33}{307}{51.97}{83.07}{308}
\emline{51.97}{83.07}{309}{54.72}{82.68}{310}
\emline{54.72}{82.68}{311}{57.23}{82.16}{312}
\emline{57.23}{82.16}{313}{59.51}{81.51}{314}
\emline{59.51}{81.51}{315}{61.56}{80.73}{316}
\emline{61.56}{80.73}{317}{63.38}{79.82}{318}
\emline{63.38}{79.82}{319}{64.97}{78.78}{320}
\emline{64.97}{78.78}{321}{66.32}{77.61}{322}
\emline{66.32}{77.61}{323}{67.45}{76.31}{324}
\emline{67.45}{76.31}{325}{68.35}{74.88}{326}
\emline{68.35}{74.88}{327}{69.01}{73.33}{328}
\emline{69.01}{73.33}{329}{69.45}{71.64}{330}
\emline{69.45}{71.64}{331}{69.67}{68.67}{332}
\emline{70.00}{69.00}{333}{69.45}{67.19}{334}
\emline{69.45}{67.19}{335}{68.68}{65.56}{336}
\emline{68.68}{65.56}{337}{67.69}{64.08}{338}
\emline{67.69}{64.08}{339}{66.48}{62.78}{340}
\emline{66.48}{62.78}{341}{65.06}{61.64}{342}
\emline{65.06}{61.64}{343}{63.42}{60.67}{344}
\emline{63.42}{60.67}{345}{61.56}{59.86}{346}
\emline{61.56}{59.86}{347}{59.48}{59.22}{348}
\emline{59.48}{59.22}{349}{57.19}{58.75}{350}
\emline{57.19}{58.75}{351}{54.68}{58.44}{352}
\emline{54.68}{58.44}{353}{51.95}{58.31}{354}
\emline{51.95}{58.31}{355}{49.00}{58.33}{356}
\put(59.67,155.67){\makebox(0,0)[cc]{$u^0$}}
\put(90.00,97.00){\makebox(0,0)[cc]{$f(V^4)$}}
\put(69.00,90.33){\vector(-2,-1){0.2}}
\emline{77.67}{95.33}{357}{69.00}{90.33}{358}
\put(15.00,119.33){\makebox(0,0)[cc]{$f(V^5)$}}
\put(40.67,112.67){\vector(2,-1){0.2}}
\emline{26.67}{120.00}{359}{40.67}{112.67}{360}
\put(86.00,57.67){\makebox(0,0)[cc]{$u^2$}}
\put(33.67,42.67){\makebox(0,0)[cc]{$u^1$}}
\end{picture}

\vspace{-1.5cm}
\caption{Spacetime $f(V^5)$ with leaf $f(V^4)$}
\end{center}
\end{figure}


Let $(y^{0},y^{1},y^{2},y^{3})$ be a coordinate system of the spacetime $V^{4}_\alpha$.
These coordinates are accosiated with the coordinates
$(x^{0},x^{1},x^{2},x^{3},x^{4})$
of the 5-dimensional
spacetime  $V^{5}$ by the formulas
$$
\left\{
\begin{array}{ll}
         x^{0}=\alpha\exp(y^{0})\\
         x^{1}=y^{0} \\
         x^{2}=y^{1} \\
         x^{3}=y^{2} \\
         x^{4}=y^{3}.
  \end{array}\right.
$$
Then induced metric of the universe $V^{4}_\alpha$
$$
g_{ik}^{(4)}(\alpha)=G_{AB}\frac{\partial x^A}{\partial y^i}
\frac{\partial x^B}{\partial y^k}
$$
in the coordinates
$y^{0},y^{1},y^{2},y^{3}$ has the form
$$
ds^{2}_\alpha=\alpha^2(1-\beta)\exp(2y^{0})(dy^{0})^{2}-(dy^{1})^{2}-
(dy^{2})^{2}-(dy^{3})^{2}.
$$
By making simple transformation
$$
y^{0^\prime}=\alpha (1-\beta)^{1/2}\exp(y^{0}), \ y^{i^\prime}=y^i, (i=1,2,3)
$$
we get
$$
ds^{2}_\alpha=(dy^{0^\prime})^{2}-(dy^{1^\prime})^{2}-
(dy^{2^\prime})^{2}-(dy^{3^\prime})^{2}.
$$
Hence universes $V^4_\alpha$ are the flat Minkowskian spacetime.

Below for simplicity instead of the cylinder $V^{5}$ we consider its
factor-space $V^{5}/\Gamma$, where $\Gamma$  is discrete group $x^{0}\to x^{0}$,
$x^{1}\to x^{1}, x^{2}\to x^{2}+d, x^{3}\to x^{3}+d, x^{4}\to x^{4}+d, $
which acts on the cylinder (d is an integer).

Topologically the factor-space $V^{5}/\Gamma$ is homeomorphic to the space
 $$
 \R^{2}\times S^{1}\times S^{1}\times S^{1}.
 $$
It means that the physical 3-dimensional space is 3-dimensional torus
$S^{1}\times S^{1}\times S^{1}$ and has  finite volume.

\section{The large fluctuations}

The amplitude of probability of transition  from universe $V^4_{\alpha_1}$ to
universe $V^4_{\alpha_2}$ is equal to
$$
<V^4_{\alpha_1}| V^4_{\alpha_2}>=\int\limits_{V^4_{\alpha_1}}^{V^4_{\alpha_2}}
\exp\left(-\frac{i}{\hbar}S\right){\cal D}[V^5], \eqno(3)
$$
$$
S= \int\limits_{V^{5}}R\sqrt{
det\Vert G_{AB}\Vert} d^5x,
$$
where Feynman path integral is taken on all five-dimensional trajectories which
connect $<V^4_{\alpha_1},g_{ik}^{(4)}(\alpha_1)> $ and
$<V^4_{\alpha_2},g_{ik}^{(4)}(\alpha_2)>$.
Among these trajectories the fluctuations of the 5-dimensional metric (1) can be
observed.
Let us consider the fluctuations $\tilde{G}_{AB}=G_{AB}+\Delta G_{AB}$
of the 5-dimensional metric (1),
for which
$$
\tilde{S}= \int\limits_{V^{5}/\Gamma}\tilde{R}
\sqrt{det\Vert \tilde{G}_{AB}\Vert} d^5x=0.
$$
We will consider the fluctuations of the form
$$
d\tilde{I^{2}}=(G_{AB}+\Delta G_{AB})dx^{A}dx^{B}=
$$
$$
=\tilde{G}_{AB}dx^{A}dx^{B}=
$$
$$
=[1+h(x^{0})](dx^{0})^2-
\beta(x^{0}-1)^{2}(dx^{1})^{2}-(dx^{2})^{2}-
(dx^{3})^{2}-
(dx^{4})^{2},
$$
where $h(x^{0})$ is an arbitrary function, such that $h(x^{0})>-1$,
$h(a)=h(b)=0$, \ $a<b$, and $h\equiv 0$ outside of $(a,b)$, i.e. interval
$(a,b)$ is an area, where the fluctuations of metric take place (see fig.3).

\begin{figure}[h]
\begin{center}
\special{em:linewidth 0.4pt}
\unitlength 0.50mm
\linethickness{0.4pt}
\begin{picture}(90.00,155.67)
\emline{38.00}{60.00}{1}{38.61}{58.37}{2}
\emline{38.61}{58.37}{3}{39.55}{56.98}{4}
\emline{39.55}{56.98}{5}{40.80}{55.82}{6}
\emline{40.80}{55.82}{7}{42.37}{54.89}{8}
\emline{42.37}{54.89}{9}{44.25}{54.19}{10}
\emline{44.25}{54.19}{11}{46.46}{53.73}{12}
\emline{46.46}{53.73}{13}{48.98}{53.50}{14}
\emline{48.98}{53.50}{15}{54.33}{53.67}{16}
\emline{53.00}{53.33}{17}{56.39}{53.56}{18}
\emline{56.39}{53.56}{19}{59.41}{53.93}{20}
\emline{59.41}{53.93}{21}{62.06}{54.45}{22}
\emline{62.06}{54.45}{23}{64.34}{55.10}{24}
\emline{64.34}{55.10}{25}{66.25}{55.90}{26}
\emline{66.25}{55.90}{27}{67.80}{56.84}{28}
\emline{67.80}{56.84}{29}{68.97}{57.92}{30}
\emline{68.97}{57.92}{31}{70.00}{59.67}{32}
\emline{38.33}{135.67}{33}{39.16}{134.21}{34}
\emline{39.16}{134.21}{35}{40.33}{133.00}{36}
\emline{40.33}{133.00}{37}{41.83}{132.04}{38}
\emline{41.83}{132.04}{39}{43.67}{131.33}{40}
\emline{43.67}{131.33}{41}{45.83}{130.87}{42}
\emline{45.83}{130.87}{43}{48.33}{130.67}{44}
\emline{48.33}{130.67}{45}{51.16}{130.71}{46}
\emline{51.16}{130.71}{47}{54.33}{131.00}{48}
\emline{52.33}{130.67}{49}{55.16}{130.54}{50}
\emline{55.16}{130.54}{51}{57.79}{130.63}{52}
\emline{57.79}{130.63}{53}{60.21}{130.93}{54}
\emline{60.21}{130.93}{55}{62.42}{131.45}{56}
\emline{62.42}{131.45}{57}{64.42}{132.19}{58}
\emline{64.42}{132.19}{59}{66.21}{133.15}{60}
\emline{66.21}{133.15}{61}{67.80}{134.32}{62}
\emline{67.80}{134.32}{63}{69.67}{136.33}{64}
\emline{38.00}{136.67}{65}{38.69}{138.07}{66}
\emline{38.69}{138.07}{67}{39.75}{139.25}{68}
\emline{39.75}{139.25}{69}{41.19}{140.23}{70}
\emline{41.19}{140.23}{71}{43.00}{141.00}{72}
\emline{43.00}{141.00}{73}{45.19}{141.56}{74}
\emline{45.19}{141.56}{75}{47.75}{141.92}{76}
\emline{47.75}{141.92}{77}{50.69}{142.06}{78}
\emline{50.69}{142.06}{79}{54.00}{142.00}{80}
\emline{53.33}{142.00}{81}{56.82}{142.07}{82}
\emline{56.82}{142.07}{83}{59.90}{141.98}{84}
\emline{59.90}{141.98}{85}{62.57}{141.73}{86}
\emline{62.57}{141.73}{87}{64.84}{141.32}{88}
\emline{64.84}{141.32}{89}{66.69}{140.76}{90}
\emline{66.69}{140.76}{91}{68.13}{140.04}{92}
\emline{68.13}{140.04}{93}{69.17}{139.16}{94}
\emline{69.17}{139.16}{95}{69.79}{138.12}{96}
\emline{69.79}{138.12}{97}{70.00}{136.67}{98}
\emline{38.00}{71.33}{99}{39.28}{69.44}{100}
\emline{39.28}{69.44}{101}{40.72}{67.86}{102}
\emline{40.72}{67.86}{103}{42.29}{66.59}{104}
\emline{42.29}{66.59}{105}{44.02}{65.63}{106}
\emline{44.02}{65.63}{107}{45.89}{64.98}{108}
\emline{45.89}{64.98}{109}{47.90}{64.64}{110}
\emline{47.90}{64.64}{111}{50.07}{64.60}{112}
\emline{50.07}{64.60}{113}{54.33}{65.33}{114}
\emline{53.67}{65.00}{115}{55.82}{65.72}{116}
\emline{55.82}{65.72}{117}{57.81}{66.57}{118}
\emline{57.81}{66.57}{119}{59.66}{67.57}{120}
\emline{59.66}{67.57}{121}{61.35}{68.70}{122}
\emline{61.35}{68.70}{123}{62.89}{69.97}{124}
\emline{62.89}{69.97}{125}{64.28}{71.38}{126}
\emline{64.28}{71.38}{127}{65.52}{72.93}{128}
\emline{65.52}{72.93}{129}{66.61}{74.61}{130}
\emline{66.61}{74.61}{131}{67.55}{76.44}{132}
\emline{67.55}{76.44}{133}{68.34}{78.40}{134}
\emline{68.34}{78.40}{135}{68.97}{80.50}{136}
\emline{68.97}{80.50}{137}{69.45}{82.73}{138}
\emline{69.45}{82.73}{139}{69.79}{85.11}{140}
\emline{69.79}{85.11}{141}{70.00}{88.67}{142}
\emline{38.00}{78.00}{143}{38.26}{79.63}{144}
\emline{38.26}{79.63}{145}{39.04}{80.96}{146}
\emline{39.04}{80.96}{147}{40.34}{82.00}{148}
\emline{40.34}{82.00}{149}{42.17}{82.74}{150}
\emline{42.17}{82.74}{151}{44.51}{83.19}{152}
\emline{44.51}{83.19}{153}{48.67}{83.33}{154}
\emline{69.33}{62.33}{155}{68.75}{61.04}{156}
\emline{68.75}{61.04}{157}{67.87}{59.88}{158}
\emline{67.87}{59.88}{159}{66.70}{58.85}{160}
\emline{66.70}{58.85}{161}{65.23}{57.95}{162}
\emline{65.23}{57.95}{163}{63.46}{57.18}{164}
\emline{63.46}{57.18}{165}{61.39}{56.55}{166}
\emline{61.39}{56.55}{167}{59.03}{56.05}{168}
\emline{59.03}{56.05}{169}{56.36}{55.67}{170}
\emline{56.36}{55.67}{171}{53.41}{55.43}{172}
\emline{53.41}{55.43}{173}{47.33}{55.33}{174}
\emline{69.67}{84.67}{175}{69.21}{87.25}{176}
\emline{69.21}{87.25}{177}{68.57}{89.62}{178}
\emline{68.57}{89.62}{179}{67.73}{91.79}{180}
\emline{67.73}{91.79}{181}{66.71}{93.75}{182}
\emline{66.71}{93.75}{183}{65.49}{95.51}{184}
\emline{65.49}{95.51}{185}{64.09}{97.07}{186}
\emline{64.09}{97.07}{187}{62.50}{98.42}{188}
\emline{62.50}{98.42}{189}{60.72}{99.57}{190}
\emline{60.72}{99.57}{191}{58.33}{100.67}{192}
\emline{70.00}{137.33}{193}{70.00}{59.00}{194}
\emline{48.67}{73.67}{195}{50.49}{73.93}{196}
\emline{50.49}{73.93}{197}{52.23}{74.35}{198}
\emline{52.23}{74.35}{199}{53.89}{74.93}{200}
\emline{53.89}{74.93}{201}{55.46}{75.67}{202}
\emline{55.46}{75.67}{203}{56.95}{76.57}{204}
\emline{56.95}{76.57}{205}{58.36}{77.64}{206}
\emline{58.36}{77.64}{207}{59.69}{78.87}{208}
\emline{59.69}{78.87}{209}{60.93}{80.25}{210}
\emline{60.93}{80.25}{211}{62.09}{81.80}{212}
\emline{62.09}{81.80}{213}{63.17}{83.52}{214}
\emline{63.17}{83.52}{215}{64.17}{85.39}{216}
\emline{64.17}{85.39}{217}{65.08}{87.42}{218}
\emline{65.08}{87.42}{219}{65.91}{89.62}{220}
\emline{65.91}{89.62}{221}{66.66}{91.97}{222}
\emline{66.66}{91.97}{223}{67.32}{94.49}{224}
\emline{67.32}{94.49}{225}{67.90}{97.17}{226}
\emline{67.90}{97.17}{227}{68.40}{100.01}{228}
\emline{68.40}{100.01}{229}{68.82}{103.01}{230}
\emline{68.82}{103.01}{231}{69.15}{106.18}{232}
\emline{69.15}{106.18}{233}{69.40}{109.50}{234}
\emline{69.40}{109.50}{235}{69.57}{112.99}{236}
\emline{69.57}{112.99}{237}{69.67}{118.67}{238}
\emline{69.67}{118.33}{239}{69.09}{120.73}{240}
\emline{69.09}{120.73}{241}{68.18}{122.73}{242}
\emline{68.18}{122.73}{243}{66.95}{124.32}{244}
\emline{66.95}{124.32}{245}{65.39}{125.50}{246}
\emline{65.39}{125.50}{247}{63.51}{126.27}{248}
\emline{63.51}{126.27}{249}{60.33}{126.67}{250}
\put(54.33,148.00){\vector(0,1){0.2}}
\emline{54.33}{59.33}{251}{54.33}{148.00}{252}
\put(81.67,59.67){\vector(1,0){0.2}}
\emline{54.00}{59.67}{253}{81.67}{59.67}{254}
\put(38.67,48.67){\vector(-4,-3){0.2}}
\emline{54.00}{59.67}{255}{38.67}{48.67}{256}
\emline{38.00}{136.00}{257}{38.00}{60.00}{258}
\emline{38.00}{79.33}{259}{39.21}{77.39}{260}
\emline{39.21}{77.39}{261}{40.70}{75.85}{262}
\emline{40.70}{75.85}{263}{42.46}{74.72}{264}
\emline{42.46}{74.72}{265}{44.49}{73.99}{266}
\emline{44.49}{73.99}{267}{48.33}{73.67}{268}
\emline{38.00}{71.00}{269}{38.40}{72.68}{270}
\emline{38.40}{72.68}{271}{39.20}{74.00}{272}
\emline{39.20}{74.00}{273}{40.41}{74.96}{274}
\emline{40.41}{74.96}{275}{42.02}{75.56}{276}
\emline{42.02}{75.56}{277}{44.03}{75.81}{278}
\emline{44.03}{75.81}{279}{48.67}{75.33}{280}
\put(69.67,62.33){\vector(1,-4){0.2}}
\emline{47.67}{75.67}{281}{50.65}{75.51}{282}
\emline{50.65}{75.51}{283}{53.41}{75.21}{284}
\emline{53.41}{75.21}{285}{55.96}{74.79}{286}
\emline{55.96}{74.79}{287}{58.29}{74.24}{288}
\emline{58.29}{74.24}{289}{60.41}{73.56}{290}
\emline{60.41}{73.56}{291}{62.31}{72.74}{292}
\emline{62.31}{72.74}{293}{63.99}{71.80}{294}
\emline{63.99}{71.80}{295}{65.46}{70.73}{296}
\emline{65.46}{70.73}{297}{66.71}{69.53}{298}
\emline{66.71}{69.53}{299}{67.74}{68.20}{300}
\emline{67.74}{68.20}{301}{68.56}{66.74}{302}
\emline{68.56}{66.74}{303}{69.16}{65.15}{304}
\emline{69.16}{65.15}{305}{69.67}{62.33}{306}
\emline{49.00}{83.33}{307}{51.97}{83.07}{308}
\emline{51.97}{83.07}{309}{54.72}{82.68}{310}
\emline{54.72}{82.68}{311}{57.23}{82.16}{312}
\emline{57.23}{82.16}{313}{59.51}{81.51}{314}
\emline{59.51}{81.51}{315}{61.56}{80.73}{316}
\emline{61.56}{80.73}{317}{63.38}{79.82}{318}
\emline{63.38}{79.82}{319}{64.97}{78.78}{320}
\emline{64.97}{78.78}{321}{66.33}{77.61}{322}
\emline{66.33}{77.61}{323}{67.45}{76.31}{324}
\emline{67.45}{76.31}{325}{68.35}{74.89}{326}
\emline{68.35}{74.89}{327}{69.02}{73.33}{328}
\emline{69.02}{73.33}{329}{69.45}{71.64}{330}
\emline{69.45}{71.64}{331}{69.67}{68.67}{332}
\emline{70.00}{69.00}{333}{69.45}{67.19}{334}
\emline{69.45}{67.19}{335}{68.68}{65.56}{336}
\emline{68.68}{65.56}{337}{67.69}{64.08}{338}
\emline{67.69}{64.08}{339}{66.48}{62.78}{340}
\emline{66.48}{62.78}{341}{65.06}{61.64}{342}
\emline{65.06}{61.64}{343}{63.42}{60.67}{344}
\emline{63.42}{60.67}{345}{61.56}{59.86}{346}
\emline{61.56}{59.86}{347}{59.48}{59.22}{348}
\emline{59.48}{59.22}{349}{57.19}{58.75}{350}
\emline{57.19}{58.75}{351}{54.68}{58.44}{352}
\emline{54.68}{58.44}{353}{51.95}{58.30}{354}
\emline{51.95}{58.30}{355}{49.00}{58.33}{356}
\put(59.67,155.67){\makebox(0,0)[cc]{$u^0$}}
\put(40.67,112.67){\vector(2,-1){0.2}}
\emline{26.67}{120.00}{359}{40.67}{112.67}{360}
\put(86.00,57.67){\makebox(0,0)[cc]{$u^2$}}
\put(33.67,42.67){\makebox(0,0)[cc]{$u^1$}}
\emline{38.00}{86.00}{361}{38.82}{84.45}{362}
\emline{38.82}{84.45}{363}{39.91}{83.15}{364}
\emline{39.91}{83.15}{365}{41.29}{82.08}{366}
\emline{41.29}{82.08}{367}{42.93}{81.26}{368}
\emline{42.93}{81.26}{369}{44.85}{80.68}{370}
\emline{44.85}{80.68}{371}{47.05}{80.34}{372}
\emline{47.05}{80.34}{373}{49.52}{80.24}{374}
\emline{49.52}{80.24}{375}{54.67}{80.67}{376}
\emline{38.00}{96.67}{377}{38.87}{94.72}{378}
\emline{38.87}{94.72}{379}{39.93}{93.06}{380}
\emline{39.93}{93.06}{381}{41.20}{91.69}{382}
\emline{41.20}{91.69}{383}{42.68}{90.59}{384}
\emline{42.68}{90.59}{385}{44.35}{89.78}{386}
\emline{44.35}{89.78}{387}{46.23}{89.24}{388}
\emline{46.23}{89.24}{389}{48.31}{89.00}{390}
\emline{48.31}{89.00}{391}{50.59}{89.03}{392}
\emline{50.59}{89.03}{393}{54.67}{89.67}{394}
\emline{54.67}{80.67}{395}{58.16}{80.84}{396}
\emline{58.16}{80.84}{397}{61.19}{81.18}{398}
\emline{61.19}{81.18}{399}{63.77}{81.67}{400}
\emline{63.77}{81.67}{401}{65.88}{82.33}{402}
\emline{65.88}{82.33}{403}{67.53}{83.14}{404}
\emline{67.53}{83.14}{405}{69.33}{85.00}{406}
\emline{54.00}{89.33}{407}{57.02}{89.43}{408}
\emline{57.02}{89.43}{409}{59.75}{89.73}{410}
\emline{59.75}{89.73}{411}{62.19}{90.22}{412}
\emline{62.19}{90.22}{413}{64.33}{90.92}{414}
\emline{64.33}{90.92}{415}{66.19}{91.81}{416}
\emline{66.19}{91.81}{417}{67.75}{92.90}{418}
\emline{67.75}{92.90}{419}{69.02}{94.18}{420}
\emline{69.02}{94.18}{421}{70.00}{95.67}{422}
\emline{40.00}{92.67}{423}{40.00}{83.00}{424}
\emline{42.67}{90.67}{425}{42.67}{81.33}{426}
\emline{45.33}{89.33}{427}{45.33}{80.67}{428}
\emline{48.33}{89.00}{429}{48.00}{80.33}{430}
\emline{51.00}{89.00}{431}{51.00}{79.67}{432}
\put(56.33,89.33){\rule{0.67\unitlength}{-8.67\unitlength}}
\emline{57.00}{89.33}{433}{57.00}{81.00}{434}
\emline{59.33}{89.67}{435}{59.33}{81.00}{436}
\emline{62.67}{90.00}{437}{62.67}{81.33}{438}
\put(65.00,91.00){\rule{-0.33\unitlength}{-8.67\unitlength}}
\emline{64.67}{90.67}{439}{64.67}{82.00}{440}
\emline{66.67}{91.67}{441}{67.00}{83.33}{442}
\put(82.33,97.00){\makebox(0,0)[cc]{$b-1$}}
\put(82.00,82.33){\makebox(0,0)[cc]{$a-1$}}
\put(16.67,120.33){\makebox(0,0)[cc]{$f(V^5)$}}
{\footnotesize
\put(17.67,90.33){\makebox(0,0)[cc]{Zone of}}
\put(17.67,85.33){\makebox(0,0)[cc]{fluctuations}}
}
\put(41.67,88.00){\vector(1,0){0.2}}
\emline{28.67}{89.00}{443}{41.67}{88.00}{444}
\end{picture}

\vspace{-1.5cm}

\caption{Zone of fluctuations of $V^5$}
\end{center}

\end{figure}
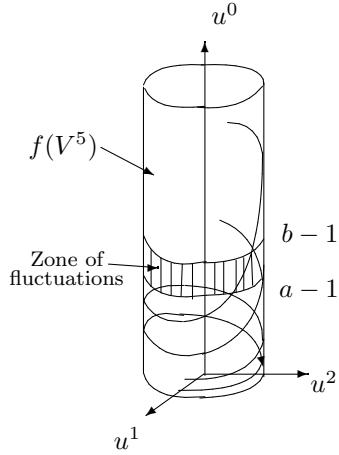

We have
$$
\tilde{R}_{00}=\frac{1}{2(1+h)(x^{0}-1)}\frac{dh}{dx^{0}}, \
\tilde{R}_{11}=-\frac{\beta(x^0-1)}{2(1+h)^2}\frac{dh}{dx^{0}}, \
$$
the scalar curvature is
$$
\tilde{R}=
\frac{1}{(1+h)^{2}(x^{0}-1)}\frac{dh}{dx^{0}},
$$
and the determinant of the metrical tensor is
$$
 det\Vert \tilde{G}_{AB}\Vert=(1+h)\beta(x^{0}-1)^{2}.
$$
So
$$
\tilde{S}= \int\limits_{V^{5}/\Gamma} \beta^{1/2}\frac{1}{(1+h)^{3/2}}\frac{dh}
{dx^{0}}d^5x=
$$
$$
=\int\limits^{b}_{a}\frac{\beta^{1/2}}{(1+h)^{3/2}}\frac{dh}{dx^{0}}dx^{0}
\int\limits_{\R^3/\Gamma}dx^1x^2dx^3.
$$
The integration over $x^{1}, x^{2}, x^{3}, x^{4}$ gives a constant. Let's
consider the integral over $x^{0}$
$$
\int\limits^{b}_{a}\frac{1}{(1+h)^{3/2}}dh=
-\frac{2}{(1+h)^{1/2}}|^b_a=0,
$$
because $h(a)=h(b)$. Then the contribution of such fluctuations in
the path integral  over 5-dimensional trajectories is equal to contribution
of basic spacetime $V^5$. In other words the our basic spacetime can
have large fluctuations, which change physical properties of the universe
$V^4_\alpha$.

Indeed, under such fluctuations the 4-dimensional metric $g_{ik}^{(4)}(\alpha)$
takes the form
$$
d\tilde{s}^{2}=\alpha^2(1-\beta+h)\exp(2y^{0})(dy^{0})^{2}-
(dy^{1})^{2} -(dy^{2})^{2}-(dy^{3})^{2}.
$$
The geometry of this spacetime is still flat. But the component
$g_{00}^{(4)}(\alpha)$ of the
metric tensor of the universe $V^4_\alpha$ has been changed. This change
means the change of the gauge on the axis $y^{0}$ and, therefore, the
change of the speed of light. It occurs instantaneously under all 3-dimensional
physical space and, hense, cannot be observed inside the universe $V^{4}_\alpha$.

An interesting phenomenon occurs when $h\to 1$. Herewith the signanture
of the spacetime
$V^{4}_\alpha$ is changed from $(+ - - -)$ to $(- - - -)$. The new signature
$(- - - -)$
means the cessation of all physical processes \cite{3}. The universe
congeals for the
arbitrary period of time. This situation can occur in any moment and
is not observed.

Note that the  stress-energy tensor of our fluctuations
$$
\tilde{T}_{22}=\tilde{T}_{33}=\tilde{T}_{44}=-\frac{1}{(1+h)^{2}(x^{0}-1)}\frac{dh}{dx^{0}}
$$
is not physical one.

\section{Choosing of the function $h(x^{0})$}

There are many functions $h(x^{0})$ which satisfy the condition
$h(x^0)>-1$, $h(a)=h(b)=0$ $(a>b)$. Here are some examples:
$$
 h(x^0)=C\sin(\frac{2\pi x^0}{a})\sin(\frac{2\pi x^0}{b}),
$$
where $C$ is a constant or a function $C(x^0)$ such as $0<1-\beta<C$.
Over this function the metric of $V^4$ changes the signature when
$h(x^0)<-1+\beta $.
$$
 h(x^0)=(x^0-a)^{2}(x^0-b)^2
$$
With this function the metric of $V^4$ doesn't change the signature.
But if $h(x^0)$ is
$$
 h(x^0)=C(x^0)(x^0-a)^{2}(x^0-b)^2, \ \ \
(-1<C(x^0)),
$$
 the metric can change the signature if function $C(x^0)$ is suitable.

\end{document}